\documentclass[a4paper,fleqn]{cas-dc}

\usepackage[numbers]{natbib}
\usepackage{amssymb,amsfonts}
\usepackage{amsmath}
\DeclareMathOperator*{\argmax}{arg\,max}

\usepackage{bbm}

\usepackage{algorithm}
\usepackage{algorithmic}
\usepackage{graphicx}
\usepackage{booktabs}
\usepackage{subfigure}
\usepackage{multirow}
\usepackage{textcomp}
\usepackage{hyperref}

\begin{document}
\let\WriteBookmarks\relax
\def\floatpagepagefraction{1}
\def\textpagefraction{.001}
\shorttitle{Robust Screening of COVID-19 from Chest X-ray via Discriminative Cost-Sensitive Learning}
\shortauthors{Tianyang Li et~al.}

\title [mode = title]{Robust Screening of COVID-19 from Chest X-ray via Discriminative Cost-Sensitive Learning}                      

\author[1,2]{Tianyang Li}[style=chinese]
\cormark[1]
\address[1]{Center for Medical Artificial Intelligence, Shandong University of Traditional Chinese Medicine, Qingdao 266112, China}
\address[2]{Qingdao Academy of Chinese Medical Sciences, Shandong University of Traditional Chinese Medicine, Qingdao 266112, China}

\author[3]{Zhongyi Han}[style=chinese]
\cormark[1]
\address[3]{School of Software, Shandong University, Jinan 250101, China}

\author[1,2]{Benzheng Wei}[style=chinese,orcid=0000-0001-9640-4947]
\cormark[2]
\ead{wbz99@sina.com}

\author[4]{Yuanjie Zheng}[style=chinese]
\address[4]{School of Information Science and Engineering, Shandong Normal University, Jinan 250358, China}

\author[1]{Yanfei Hong}

\author[1,2]{Jinyu Cong}

\author[5]{Wei Zhang}

\author[5]{Xue Zhu}
\address[5]{Department of Pulmonary Disease, Affiliated Hospital of Shandong University of Traditional Chinese Medicine, Jinan 250014, China}

\author[5]{Xuxiang Lu}

\author[6]{Haifeng Wei}
\address[6]{Department of Medical Imaging, Affiliated Hospital of Shandong University of Traditional Chinese Medicine, Jinan 250014, China}

\cortext[cor1]{These authors contributed equally to this work.}
\cortext[cor2]{Corresponding author.}

\begin{abstract}
This paper addresses the new problem of automated screening of coronavirus disease 2019 (COVID-19) based on chest X-rays, which is urgently demanded toward fast stopping the pandemic. However, robust and accurate screening of COVID-19 from chest X-rays is still a globally recognized challenge because of two bottlenecks: 1) imaging features of COVID-19 share some similarities with other pneumonia on chest X-rays, and 2) the misdiagnosis rate of COVID-19 is very high, and the misdiagnosis cost is expensive. While a few pioneering works have made much progress, they underestimate both crucial bottlenecks. In this paper, we report our solution, discriminative cost-sensitive learning (DCSL), which should be the choice if the clinical needs the assisted screening of COVID-19 from chest X-rays. DCSL combines both advantages from fine-grained classification and cost-sensitive learning. Firstly, DCSL develops a conditional center loss that learns deep discriminative representation. Secondly, DCSL establishes score-level cost-sensitive learning that can adaptively enlarge the cost of misclassifying COVID-19 examples into other classes. DCSL is so flexible that it can apply in any deep neural network. We collected a large-scale multi-class dataset comprised of 2,239 chest X-ray examples: 239 examples from confirmed COVID-19 cases, 1,000 examples with confirmed bacterial or viral pneumonia cases, and 1,000 examples of healthy people. Extensive experiments on the three-class classification show that our algorithm remarkably outperforms state-of-the-art algorithms. It achieves an accuracy of 97.01\%, a precision of 97\%, a sensitivity of 97.09\%, and an F1-score of 96.98\%. These results endow our algorithm as an efficient tool for the fast large-scale screening of COVID-19.
\end{abstract}


\maketitle

\section{Introduction}
As COVID-19 continues to affect our world, automated screening of coronavirus disease 2019 (COVID-19) is urgently needed to realize large-scale screening to combat it. Since the outbreak of COVID-19 in December 2019 to date, more than 5,083,278 people have been infected around the world. And more than 329,668 deaths from the virus have been recorded according to the World Health Organization. Fast and large-scale screening is necessary to cut off the source of infection. However, the rapidly growing amount of COVID-19 cases makes global medical resources unbearable. Automated screening systems can correspondingly assist in speeding up screening and would reduce the workload of radiologists. Therefore, it is urgent to realize the automated screening of COVID-19 that helps to accelerate the large-scale screening and alleviate the global shortage of medical resources.

Nowadays, medical imaging examinations, such as chest CT, X-ray, play an essential role in the diagnosis process of COVID-19. Clinically, although nucleic acid detection is the gold standard, the availability, stability, and reproducibility of nucleic acid detection kits are problematic~\cite{wang2020deep}. For example, the quantity of nucleic acid kits is limited in many countries or regions, resulting in the slower screening of new coronary pneumonia~\cite{zhang2020covid}. Many patients with new coronary pneumonia cannot be tested in time and thus cannot be admitted to the hospital, which accelerates the widespread of the novel virus. On the contrary, medical imaging examinations can help clinical to carry out disease detection conveniently and quickly, and thus make patients be treated timely. Accordingly, medical imaging examinations with symptom observation are widely used for the early diagnosis of COVID-19 worldwide~\cite{jin2020rapid}.

X-rays have unique advantages of light, quick, and availability in the screening of COVID-19. On the one hand, regular X-ray machines can be accessed in most primary hospitals where CT scanners are insufficient. Most ambulatory care facilities have deployed X-ray units as basic diagnostic imaging. Most importantly, the easily accessible X-ray examination is beneficial for fast large-scale screening. On the other hand, the radiation dose of X-ray is a few hundredths of the chest CT~\cite{furlow2010radiation,huda2007radiation}. One statistic from the San Francisco bay area shows that hospitals add one more case of cancer for every 400 to 2,000 additional routine chest CT examinations~\cite{storrs2013ct}. This statistic indicates that a single chest CT examination increases the lifetime risk of cancer by 0.05\% - 0.25\%. Also, the X-ray examination is more economical than CT. 

Therefore, this paper addresses the novel task of automated screening of COVID-19 from chest X-rays. However, robust and accurate screening has two crucial bottlenecks. Firstly, the imaging features of COVID-19 cases share some similarities with other pneumonia cases on chest X-rays. Even radiologists cannot distinguish them based on chest CT accurately without other inspection methods~\cite{wang2020deep}, let alone based on chest X-rays. The second bottleneck is that the misdiagnosis rate of COVID-19 is very high. High misdiagnosis rate of COVID-19 has a prohibitive cost that is not only delaying the timely treatment of patients but also causing the widespread of the virus with high social costs. Misdiagnosis cost sensitivity should be fully considered in the automated screening of COVID-19. Therefore, these limitations impedes the accurate screening of COVID-19 patients from the susceptible population.

While a few pioneering works have made much progress, they neglect both crucial bottlenecks. All of them adopted common machine learning classifiers. For example, both Hassanien \emph{et al.}~\cite{hassanien2020automatic} and Sethy \emph{et al.}~\cite{sethy2020detection} used support vector machines (SVM) to screen COVID-19 based on extracted features. Farooq \emph{et al.}~\cite{farooq2020covid}, Narin \emph{et al.}~\cite{narin2020automatic}, Wang \emph{et al.}~\cite{wang2020covid} and etc adopted popular deep neural networks with a minor modification of network architecture. More interestingly, Zhang \emph{et al.}~\cite{zhang2020covid} viewed this screening task as an anomaly detection problem and proposed to use existing anomaly detection techniques. However, in fact, the screening of COVID-19 is a fine-grained cost-sensitive classification problem, as mentioned before. Only from this perspective can we design a satisfactory solution. Therefore, in this study, we attempt to design an efficient solution to combat the bottlenecks.

In this paper, we propose an innovative discriminative cost-sensitive learning (DCSL) for the robust screening of COVID-19 from chest X-rays. DCSL combines both advances from fine-grained classification and cost-sensitive learning techniques. To overcome the subtle difference bottleneck, we propose a conditional center loss function to learn the discriminative representation between fine-grained classes. By combing with vanilla loss function, the conditional center loss can discover a weighted center for each class and efficiently enlarge the inter-class manifold distance as well as enhancing the intra-class compactness in deep representation space. To combat the cost sensitivity, we propose score-level cost-sensitive learning. It introduces a score-level cost matrix to reshape the classifier confidences by modifying the classifier output, such that the COVID-19 examples have the maximum score, and the other classes have a substantially lower score. Based on the domain knowledge that the costs between misclassifying COVID-19 into different classes or misclassifying other classes into COVID-19 are not equal, we define new score-level costs to encourage the correct classification of COVID-19 class. We combine both advances into a deep neural network with end-to-end optimization, successfully achieving fine-grained cost-sensitive screening of COVID-19. A series of experiments show that our algorithm remarkably outperforms previous methods.

The contributions of this work include:
\begin{itemize}
	\item For the first time, we formulate the task of screening of COVID-19 from chest X-ray as a fine-grained cost-sensitive classification problem. Accordingly, we propose a practical solution, discriminative cost-sensitive learning, that achieves much high screening accuracy.
	\item We propose a new conditional center loss that considers the class-conditional information when learning the center points per class. The conditional center loss successfully overcomes the bottleneck of feature similarities.
	\item We propose a new score-level cost-sensitive learning that introduces a domain knowledge-based cost matrix to enlarge the cost of misclassifying COVID-19 examples into other classes. It greatly reduces the misdiagnosis rate.
\end{itemize}

The remainder of this paper is organized as follows. Section~\ref{RW} presents the related works in terms of artificial intelligence assisted analysis of COVID-19 and involved methodologies. Section~\ref{Methodology} gives in detail the proposed discriminative cost-sensitive learning. Section~\ref{Experiments} presents detailed descriptions of collected datasets, experiment settings, and exhaustive results. Section~\ref{Conclusion} concludes this work comprehensively.

\section{Related Works}
\label{RW}

This section shows related works about the automated screening of COVID-19 and involved algorithms of our work.

\subsection{Automated Medical Image Analysis about COVID-19}

To take part in the global fight against COVID-19, many studies have designed AI-empowered technologies for improving the clinical diagnosis efficiency. Shi \emph{et al.}~\cite{shi2020review} comprehensively summarized lots of emerging works, including automated screening~\cite{wang2020deep,xu2020deep,shi2020large,jin2020ai,song2020deep,gozes2020rapid,gozes2020coronavirus,jin2020development,zheng2020deep}, patient severity assessment~\cite{huang2020serial}, infection quantification~\cite{shan+2020lung}, and infection area segmentation~\cite{chen2020deep,gozes2020rapid}. Among them, automated screening received the most attention, involving chest X-ray based and chest CT based works. Since 3D CT scans have spatial complexity, existing CT based works have proposed to design three types of solutions, including patch-based methods~\cite{wang2020deep,xu2020deep,shi2020large,jin2020ai}, slice-based methods~\cite{song2020deep,gozes2020rapid,gozes2020coronavirus,jin2020development}, and 3D CT-based method~\cite{zheng2020deep}. As we mentioned before, CT has some disadvantages of expensive, high radiation, and inaccessibility, thus lots of screening works are based on 2D chest X-rays.

We comprehensively review chest X-rays based methods as follows. Hassanien \emph{et al.}~\cite{hassanien2020automatic} used a multi-level threshold segmentation algorithm to crop lung areas and adopted SVM to classify COVID-19 and normal cases based on 40 chest X-rays. Ozturk \emph{et al.}~\cite{ozturk2020classification} ensembled several feature extraction algorithms and used a stacked autoencoder with principal component analysis to make decisions. They showed that handcrafted features based classifiers perform better than deep models on small data. Several studies applied popular deep learning techniques for the screening of COVID-19. Hemdan \emph{et al.}~\cite{hemdan2020covidx} validated multiple popular deep models and demonstrated their effectiveness on this new task. Farooq \emph{et al.}~\cite{farooq2020covid} utilized the residual networks~(ResNet) to validate the screening performance of COVID-19. Narin \emph{et al.}~\cite{narin2020automatic} tested an inception architecture (Inceptionv3) on the screening task. They showed that pre-trained models are useful. The VGG16 and VGG19 networks are adopted by Hall \emph{et al.}~\cite{hall2020finding} and Apostolopoulos \emph{et al.}~\cite{apostolopoulos2020covid}, respectively. Khalifa \emph{et al.}~\cite{khalifa2020detection} used generative adversarial networks and a fine-tuned deep transfer learning model, achieving promising and effective performance. Their results also confirm that chest X-rays based screening of COVID-19 has great research significance.

Moreover, several studies designed specialized solutions for the screening of COVID-19 according to the characteristics of the task. Afshar \emph{et al.}~\cite{afshar2020covid} adopted a capsule network for handling small data. Abbas \emph{et al.}~\cite{abbas2020classification} designed a Decompose, Transfer, and Compose (DeTraC) network based on class decomposition for enhancing low variance classifiers and facilitating more flexibility to their decision boundaries. Wang \emph{et al.}~\cite{wang2020deep} proposed a new deep network called COVID-Net, which consists of stacked residual blocks for achieving easily training and deepening the architectures. To improve diagnostic performance, Ghosha \emph{et al.}~\cite{ghoshal2020estimating} used Bayesian convolutional neural networks to estimate uncertainty. More interestingly, Zhang \emph{et al.}~\cite{zhang2020covid} viewed this screening task as an anomaly detection problem and proposed to use existing anomaly detection techniques. Specifically, they used a hybrid loss that combines a binary cross-entropy loss and a deviation loss to assign anomaly scores to COVID-19 examples. While a few pioneering works have made great progress, they neglect both cost sensitivity and fine-grained bottlenecks. To the best of our knowledge, it is the first time that we insightfully view the screening of COVID-19 from chest X-rays as a fine-grained cost-sensitive classification problem.

\subsection{Involved Methods of Our Work}

\subsubsection{Fine-Grained Classification}

The goal of fine-grained classification is to classify data belonging to multiple subordinate categories, \emph{e.g.,} COVID-19, common pneumonia, normal chest X-rays. The facing problem of fine-grained classification is that these subordinate categories naturally exist small inter-class variations and large intra-class variations. The common solutions of this problem can be organized into three main paradigms, including 1) localization-classification networks based methods~\cite{zhang2014part,lin2015deep,wei2018mask}, 2) external information-based methods~\cite{sun2019learning,xu2018fine,deng2015leveraging}, and 3) end-to-end feature coding-based methods~\cite{lin2015bilinear,dubey2018maximum}. Localization-classification networks based methods first learn part-based detectors or segmentation model to localize salient parts for improving the final recognition accuracy. However, this type of paradigm needs additional part annotations. External information-based methods leverage external information, \emph{i.e.,} web data~\cite{sun2019learning}, multi-modality~\cite{xu2018fine}, or human-computer interactions~\cite{deng2015leveraging}. 

Different from the previous two paradigms, end-to-end feature coding-based methods learn a more discriminative feature representation directly~\cite{ge2019pv}. Among them, several works specifically designed extra useful loss functions for learning discriminative fine-grained representations. For instance, the contrastive loss is designed for dealing with the relationship of paired example points effectively~\cite{sun2014deep}. Triplet loss correspondingly constructs loss functions for example triplet~\cite{schroff2015facenet}. However, contrastive loss and triplet loss are required that the number of training pairs or triplets dramatically grows, with slow convergence and instability. Center loss is proposed to solve this issue by minimizing the deep feature distances of intra-class only~\cite{wen2016discriminative}. This loss function learns a center for each class and pulls the deep features of the same class to their centers efficiently. However, center loss easily suffers from the issue of class imbalance. Therefore, in this paper, we propose to learn a new conditional center loss with joint balance optimization for the robust screening of COVID-19.

\subsubsection{Cost-Sensitive Learning}

Cost-sensitive learning is a learning method that considers the cost of misclassification, and its purpose is to minimize the total cost~\cite{ling2008cost}. In the classical machine learning setting, the costs of classification errors of different classes are equal. Unfortunately, the costs are not equal in many real-world tasks. For example, in COVID-19 screening, the cost of erroneously diagnosing a COVID-19 patient to be healthy may be much higher than that of mistakenly diagnosing a COVID-19 patient to be common pneumonia. Cost-sensitive learning is proposed to handle this problem and has attracted much attention from the machine learning and data mining communities~\cite{zhou2010multi}. Existing works on misclassification costs can be categorized into two classes, including example-dependent cost~\cite{zadrozny2001learning,zadrozny2003simple,brefeld2003support} and class-dependent cost~\cite{breiman1984classification,domingos1999metacost,maloof2003learning}. Example-dependent cost-based methods consider the misclassification cost of each example and are required example-level annotations, which are impractical in real-world tasks. Therefore, most methods are focused on class-dependent costs. Cost-sensitive learning is also suitable to handle the problem of class imbalance~\cite{khan2017cost}. In this study, we introduce a score-level cost-sensitive learning approach based on an expert-provided cost matrix to improve the screening accuracy of COVID-19 from chest X-rays.


\begin{figure*}
	\centering
	\includegraphics[width=1\textwidth]{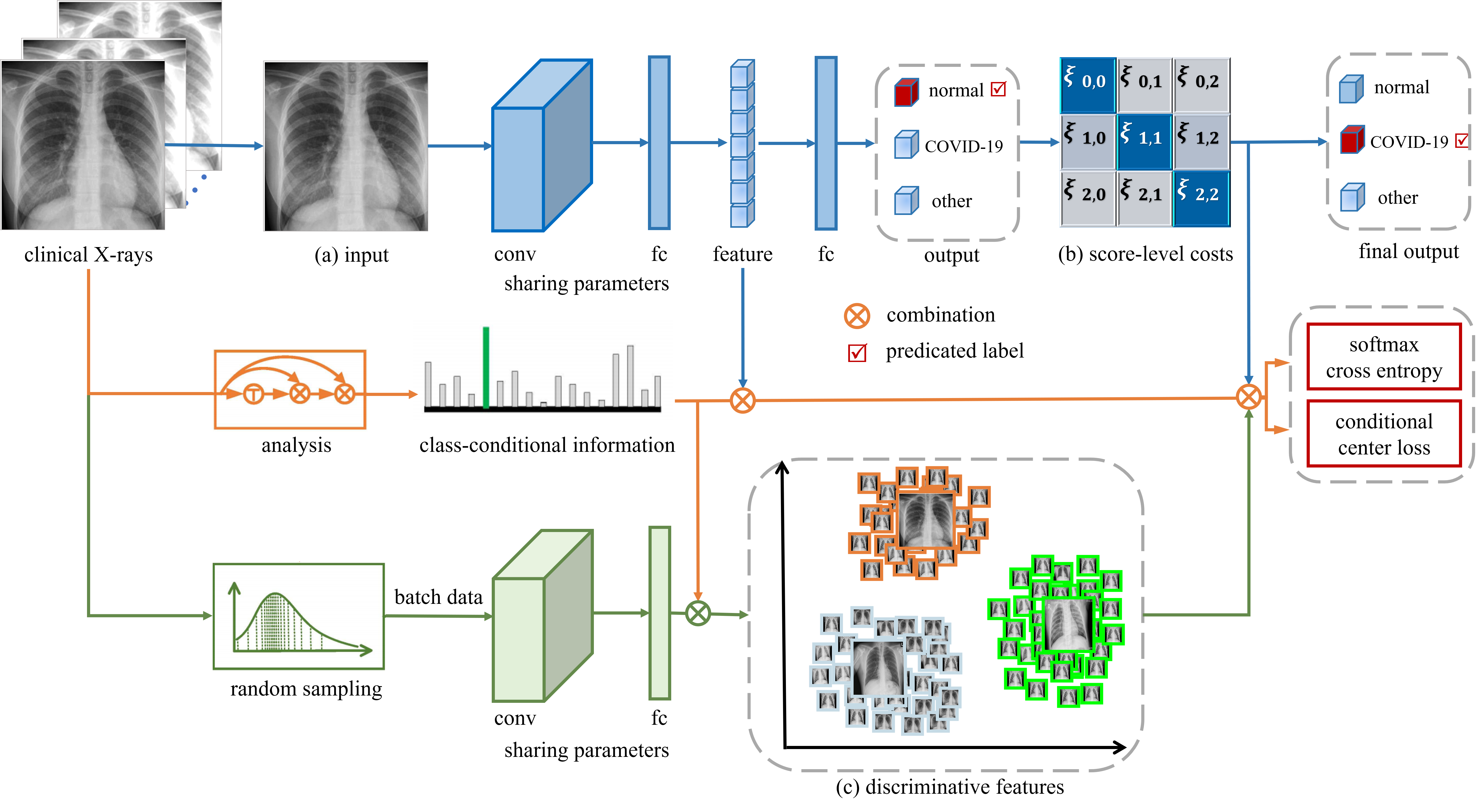}
	\caption{\label{fig:1} Overview of the proposed discriminative cost-sensitive learning (DCSL) framework. Based on a data pool of clinical X-rays, a comprehensive analysis is conducted to obtain the class-conditional information (class balance weight). In the training period, we randomly draw a batch of data to optimizer softmax cross-entropy and conditional center loss with the class-conditional information. Specifically, an input image is firstly processed by a backbone network, which mainly includes convolutional layers~(Conv) and fully-connected layers~(FC). After that, deep features are obtained to minimize conditional center loss. Score-level costs are applied to the outputs from the final output layer to get the final cost-sensitive prediction.}
\end{figure*}

\section{Methodology}
\label{Methodology}

In this section, we first introduce the necessary notations and the objective for the task of screening of COVID-19 from chest X-rays (See Section~\ref{ps}). We then present the newly-proposed discriminative cost-sensitive  learning (DCSL) framework, which consists of a conditional center loss (See Section~\ref{ccl}) and a score-level cost-sensitive learning approach (See Section~\ref{slcsl}). We finally combine the two novel modules to construct the DCSL framework for the fine-grained cost-sensitive classification problem (See Section~\ref{DCSL}).

\subsection{Problem Setting}
\label{ps}
The task of COVID-19 screening is under the familiar supervised learning setting where the learner receives a sample of $m$ labeled training examples $\{(x_i,y_i)\}_{i=1}^m$ drawn from a joint distribution $Q$ defined on $\mathcal{X} \times \mathcal{Y}$, where $\mathcal{X}$ is the example set of 2D chest X-ray images, and $\mathcal{Y}$ is the label set of patient conditions, such as COVID-19, common pneumonia, others. $\mathcal{Y}$ is $\{0, 1\}$ in binary classification and  $\{1, \dots, n\}$ in multi-class classification. Specifically, $x_i$ is any chest X-ray image of one patient, and $y_i$ is the label of this patient. We denote by $\hat{Q}$ the empirical distribution. 

We denote by $\mathcal{L}: \mathcal{Y} \times \mathcal{Y} \rightarrow \mathbb{R}$ any loss function defined over pairs of labels, such as 0-1 loss, cross-entropy loss, etc. For binary classification, we denote by $f: \mathcal{X} \rightarrow \{0,1\}$ a scoring function, which induces a labeling function $h_f: \mathcal{X} \rightarrow \mathcal{Y}$ where $h_f: x \rightarrow \argmax_{y\in \mathcal{Y}} [f(x)]_y$. For any distribution $Q$ on $\mathcal{X} \times \mathcal{Y}$ and any labeling function $h_f$, we denote $\mathcal{R}_{Q}(h_f) = \mathbbm{E}_{(x,y)\sim Q} \mathcal{L}(h_f(x),y)$ the expected risk. Our objective is to select a hypothesis $f$ out of a hypothesis set $\mathcal{F}$ with a small expected risk $\mathcal{R}_{Q}(h_f)$ on the target distribution.

\subsection{Conditional Center Loss}
\label{ccl}

This section presents the newly-proposed conditional center loss. To better understand the role of conditional center loss, we first consider the softmax based cross-entropy loss (softmax loss) that is presented as follows.
\begin{equation}
\mathcal{L}_S = -\frac{1}{m} \sum_{i=1}^{m} \log \frac{e^{W_{y_i}^{T}\boldsymbol x_i + b_{y_i}}}{\sum_{j=1}^{n}e^{W_{j}^{T}\boldsymbol x_i+b_j}}\,,
\label{softmax_loss}
\end{equation}
where $\boldsymbol x_i \in \mathbbm{R}^d$ (to reduce abuse notations) denotes the $i$th deep feature of the X-ray image $x_i$, belonging to the class $y_i$. $W_j \in \mathbbm{R}^d\ $ denotes the $j$th column of the weights $W \in \mathbbm{R}^{d\times n}$ in the last fully-connected layer, and $b \in \mathbbm{R}^{n}$ is the bias term. $m$ and $n$ are the size of mini-batch and the number of classes, respectively. Intuitively, the softmax loss first computes the probability of correct classification and then takes the logarithm of this probability. Since the logarithm value of probability is negative, a minus sign is added in front of it. While the softmax loss is good at common object recognition tasks, it cannot learn enough discriminative features in processing fine-grained classification tasks in which significant intra-class variations exist.

According to our in-depth analysis of the characteristics of the screening of COVID-19 from chest X-rays, we, with keen insight, view it as a fine-grained classification problem. As we discussed before, the center loss has advantages of flexibility, easy-to-implement, and stability. Therefore, we initially leverage the center loss to learn more discriminative features. The goal of center loss is to directly improve the intra-class compactness, which is conducive to the discriminative feature learning. Center loss is widely embedded between the fully connected layers of deep neural networks for decreasing the intra-class variations in the representation space (the dimension is two). It is commonly appeared with softmax loss and used for face recognition and fine-grained classification. The center loss function is formulated as follows.

\begin{equation}
\mathcal{L}_C = \frac{1}{m} \sum_{i=1}^{m}\parallel \boldsymbol x_i - \boldsymbol c_{y_i}\parallel_2^2\,,
\end{equation}
where the $\boldsymbol c_{y_i}$ denotes the $y_i$th class center of deep features. The center is updated according to the mini-batch data. And it is computed by averaging the features of the corresponding classes. A scalar $\alpha$ is used to control the learning rate of the centers. The update equation of $\boldsymbol c_{y_i}$ is represented as follows.

\begin{equation}
c_{y_i}^{t+1} = c_{y_i}^t - \alpha \triangle \boldsymbol c_j, \, \forall \, y_i = j\,, \, {\rm in\,\, which}\,, 
\end{equation}
\begin{equation}
\triangle \boldsymbol c_j = \frac{\sum_{i=1}^{m}\delta(y_i = j) \cdot (\boldsymbol c_j - \boldsymbol x_i)}{1 + \sum_{i=1}^{m}\delta(y_i = j)}\,,
\end{equation}
where $\delta$ is an indicator function in which $\delta = 1$ if the condition is satisfied, and  $\delta = 0$ if not. $t$ denotes the iteration of training. The final joint loss function is given by
\begin{equation}
\begin{split}
\mathcal{L}_{SC} &= \mathcal{L}_S + \lambda_C \mathcal{L}_C \\
&= \frac{1}{m}\sum_{i=1}^{m}(\log\frac{-e^{W_{y_i}^{T}\boldsymbol x_i + b_{y_i}}}{\sum_{j=1}^{n}e^{W_{j}^{T}\boldsymbol x_i+b_j}} + \lambda_C\parallel \boldsymbol x_i - \boldsymbol c_{y_i}\parallel_2^2)\,,
\end{split}
\end{equation}
where $\lambda_C$ is used for balancing the joint loss function.

While the joint loss function has achieved great success in practice, however, it quickly losses efficiency in the class imbalance situation. In other words, the center loss does not work in the screening task of COVID-19 according to our observations. After an in-depth analysis, we found that the problem is that the learned center points are unrepresentative. To handle this problem, we propose a conditional center loss that considers the class-conditional information when updating the center points and optimizing the center loss. We denote by $w_j$ the weight of $j$th class, and $w_j$ is computed by the ratio between the number of $j$th class's training examples and the total training examples. The update equation of center points is reformulated by
\begin{equation}
c_{y_i}^{t+1} = c_{y_i}^t - \alpha w_j \triangle \boldsymbol c_j, \, \forall \, y_i = j\,, \, {\rm in\,\, which}\,, 
\end{equation}

\begin{equation}
\triangle \boldsymbol c_j = \frac{\sum_{i=1}^{m} w_j \delta(y_i = j) \cdot (\boldsymbol c_j - \boldsymbol x_i)}{1 + \sum_{i=1}^{m}\delta(y_i = j)}\,,
\end{equation}

Meanwhile, we found that embedding the clas-conditional information into the center loss can significantly improve the total screening accuracy. The reason is that it makes the center loss learn more balance center points and thus enhances the intra-class compactness to obtain discriminative deep features. Accordingly, the conditional center loss with softmax loss is represented as follows.

\begin{equation}
\begin{split}
\mathcal{L}_{SC} &= \mathcal{L}_S + \lambda_C \mathcal{L}_C \\
&= \frac{1}{m} \sum_{i=1}^{m}(\log\frac{- w_j e^{W_{y_i}^{T}\boldsymbol x_i + b_{y_i}}}{\sum_{j=1}^{n}e^{W_{j}^{T}\boldsymbol x_i+b_j}} + w_j \lambda_C\parallel \boldsymbol x_i - \boldsymbol c_{y_i}\parallel_2^2)\,.
\end{split}
\end{equation}
The conditional center loss can effectively handle the problem that imaging features of COVID-19 share some similarities with other pneumonia on chest X-rays by enlarging their feature distance in high-level representation space.

\subsection{Score-Level Cost-Sensitive Learning}
\label{slcsl}
The goal of cost-sensitive learning is to classify examples from essential classes such as COVID-19 correctly. We propose a score-level cost-sensitive learning module that can efficiently learn robust feature representations for both the critical and common classes. It thus can enhance the accuracy of COVID-19 without unduly sacrificing the precision of the overall accuracy. Generally speaking, we introduce a handcrafted cost matrix whose design is based on clinical expert experience and then incorporate it after the output layer of deep neural networks. In this manner, we can directly modify the learning process to incorporate class-dependent costs during training and testing, without affecting the training and testing time of the original network. We will show that the proposed algorithm can efficiently work for the screening of COVID-19 and can be inserted into any deep neural networks. In this section, we first present the traditional cost-sensitive learning, which usually adds the cost matrix into the loss functions. We then detail the score-level cost-sensitive learning with an advanced learning strategy.

Formally, we denote by $\xi'$ the cost matrix whose diagonal $\xi_{p,p}'$ represents the benefit for a correct prediction. We also denote by $\xi_{p,q}$ the misclassification cost of classifying an example belonging to a class $p$ into a different class $q$. The expected risk defined on the target distribution $Q$ is given by 

\begin{equation}
\mathcal{R}(p|x) = \sum_q \xi_{p,q}' P(q|x)\,,
\end{equation}
where $P(q|x)$ is the posterior probability over all possible classes given an example $x$. The goal of a classifier is to minimize the expected risk $\mathcal{R}(p|x)$, however, which cannot be reached in practice. Thus, we use its empirical distribution of $\hat{Q}$ to minimize the empirical risk as follows.

\begin{equation}
\begin{split}
\hat{\mathcal{R}}(f) &=  \mathbbm{E}_{(x,y)\sim \hat{Q}} \mathcal{L}(h_f(x_i), y_i) \\ 
& = \frac{1}{m} \sum_{i=1}^m \mathcal{L}(\xi', f(x_i), z_i)\,,
\end{split}
\end{equation}
where $f(\cdot)$ denotes a neural netowrk and $f(x_i)$ is the neural netowrk output. $z_i \in \mathbbm{R}^n$ denotes the one-hot of the label $y_i$. For neural networks, loss function $\mathcal{L}$ can be a cross-entropy loss function with softmax, which is penalized by the cost matrix $\xi'$ as follows.

\begin{equation}
\mathcal{L}(\xi') = -\frac{1}{m} \sum_{i=1}^{m} \xi_{p,q}' \log \frac{e^{W_{y_p}^{T}\boldsymbol x_i + b_{y_p}}}{\sum_{p=1}^{n}e^{W_{p}^{T}\boldsymbol x_i+b_p}}\,,
\label{weighted_softmax_loss}
\end{equation}
where $\boldsymbol x$ denote the deep feature from the output of the penultimate layer. The entries of a handcrafted cost matrix usually have the form of
\begin{align}
\xi' = \begin{cases} \xi_{p,q}' = 0, & p = q\,,
\\ \xi_{p,q}' \in \mathbbm{R}, & p \neq q\,. \end{cases}
\label{assumption}
\end{align}
Inserting such a cost matrix into loss function can increase the corresponding loss value of an important class. However, such a manner would make the training process of neural networks unstable and can lead to non-convergence~\cite{khan2017cost}. Therefore, we propose an alternative cost-sensitive learning.

\begin{figure*}
	\begin{minipage}[ht]{0.248\linewidth} 
		\centering
		\includegraphics[width=0.8\textwidth, height=0.7\textwidth]{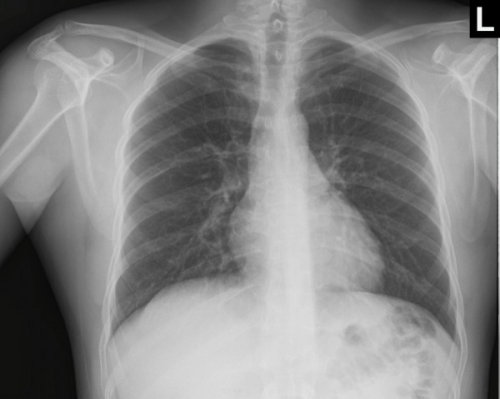}
		\centerline{(a) Normal}
		\label{fig2:side:a}
	\end{minipage}%
	\begin{minipage}[ht]{0.248\linewidth}
		\centering
		\includegraphics[width=0.8\textwidth, height=0.7\textwidth]{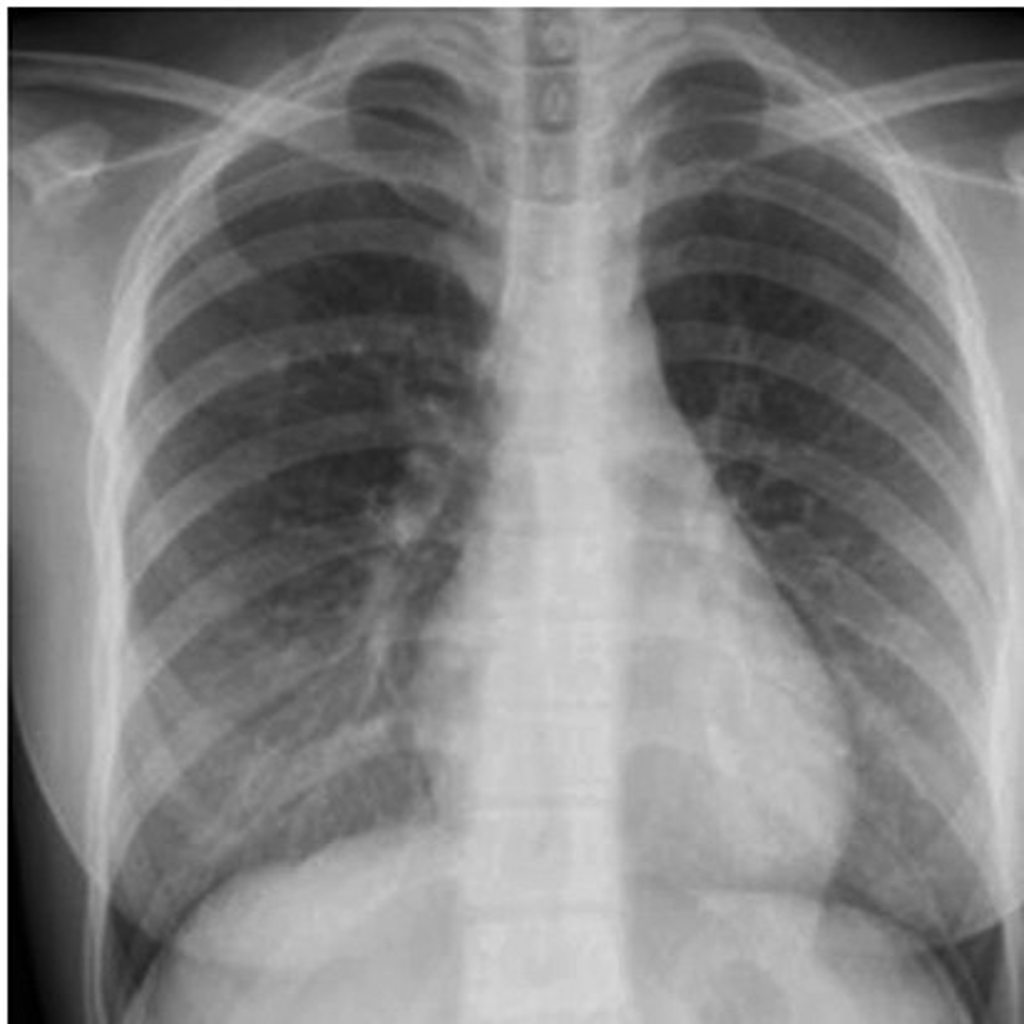}
		\centerline{(b) COVID-19}
		\label{fig2:side:b}
	\end{minipage}
	\begin{minipage}[ht]{0.248\linewidth} 
		\centering
		\includegraphics[width=0.8\textwidth, height=0.7\textwidth]{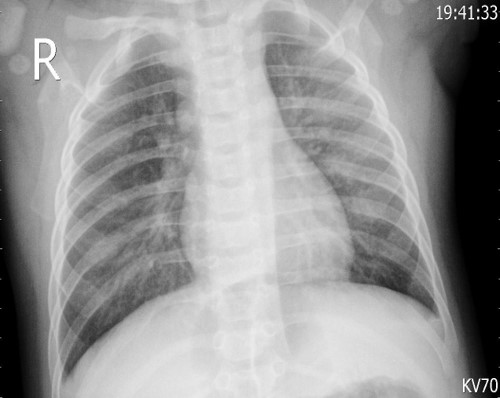}
		\centerline{(c) Viral Pneumonia}
		\label{fig2:side:c}
	\end{minipage}%
	\begin{minipage}[ht]{0.248\linewidth}
		\centering
		\includegraphics[width=0.8\textwidth, height=0.7\textwidth]{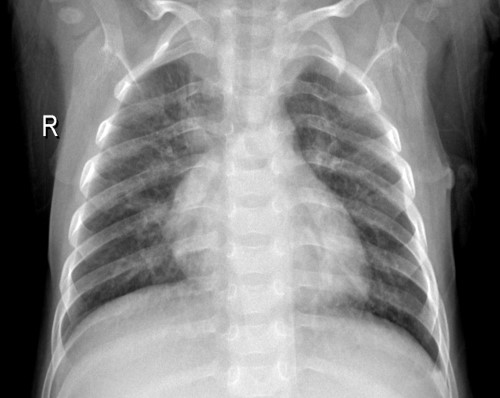}
		\centerline{(d) Bacterial Pneumonia}
		\label{fig2:side:d}
	\end{minipage}
	\caption{\label{fig:2} The visualization of representative chest X-ray images from the collected dataset. They show high similarities.}
\end{figure*}

To make the learning process more stable and convergence, we propose a new score-level cost matrix $\xi$ to modify the output of the last layer of a convolutional neural network (CNN). As shown in Figure~\ref{fig:1}, the location of the score-level cost matrix is after the output layer, and before the loss layer with softmax. We introduce the new score-level costs to encourage the correct classification of essential classes. Therefore, the CNN output $f(x_i)$ is modified using the cost matrix as follows.
\begin{equation}
o_i = \xi f(x_i)\,,
\end{equation}
where $f(x_i) \in \mathbbm{R}^{1 \times n}$ and $\xi \in \mathbbm{R}^{n \times n}$, such that $o_i \in \mathbbm{R}^{1 \times n}$. During the training process, the output weights are modified by the score-level cost matrix to reshape the classifier confidences such that the desired class has the maximum score, and the other classes have a considerably low score. Note that the score-level costs perturb the classifier confidences. Such perturbation allows the classifier to give more attention to the desired classes. In practice, all cost values in $\xi$ are positive, which enables a smooth training process. When using the score-level cost matrix, the cross-entropy loss function with softmax is finally can be revised by

\begin{equation}
\mathcal{L}(\xi) = -\frac{1}{m} \sum_{i=1}^{m} \log \frac{e^{\xi (W_{y_p}^{T}\boldsymbol x_i + b_{y_p})}}{\sum_{p=1}^{n} e^{\xi (W_{p}^{T}\boldsymbol x_i+b_p)}}\,.
\end{equation}

\begin{algorithm}[t] 
	\caption{DCSL Algorithm.} 
	\label{alg:Framwork} 
	\begin{algorithmic}[1]
		\REQUIRE ~~\\
		Training data $(x,y) \sim \hat{Q}$;\\
		Initialize parameters $\theta$ in backbone layers, parameters $W$, $b$ in the output layer, and ${\boldsymbol c_p|p = 1, 2, ..., n}$;\\
		Hyperparameter class balance weight $w$, $\xi$, $\lambda_C$ and $\alpha$;\\
		The learning rate $\eta$;\\
		The number of iteration $t$.
		\ENSURE The parameters $\theta$, $W$, $b$.\\
		\WHILE{not converge}
		\STATE $t \Leftarrow t + 1$
		\STATE \textbf{Obtain} batch $x_b$
		\STATE \textbf{Obtain} deep features $\boldsymbol x_b$
		\STATE \textbf{Obtain} outputs $o_i = \xi (W \boldsymbol x_b + b)$
		\STATE \textbf{Compute} the joint loss by $\mathcal{L}_{DCSL}^{t} ({\boldsymbol x_b}, o_i, {\boldsymbol c_p^t}, w_p,\lambda_C)$
		\STATE \textbf{Update} $c_{j}^{t+1} \Leftarrow c_{p}^{t} - \alpha w_p\cdot \triangle \boldsymbol c_{p}^{t}$
		\STATE \textbf{Update} $W \Leftarrow W - \eta \nabla \mathcal{L}_{DCSL}^t$
		\STATE \textbf{Update} $b \Leftarrow b - \eta \nabla \mathcal{L}_{DCSL}^t$
		\STATE \textbf{Update} $\theta \Leftarrow \theta - \eta \nabla \mathcal{L}_{DCSL}^t$ \\
		\ENDWHILE
	\end{algorithmic}
\end{algorithm}

\subsection{Discriminative Cost-Sensitive Learning}
\label{DCSL}
Combining the conditional center loss and the score-level cost-sensitive learning, we propose the discriminative cost-sensitive learning (DCSL) framework. As shown in Figure~\ref{fig:1}, given a chest X-ray image $x_i$, DCSL first uses a backbone network parameterized by $\theta$ to extract deep features $\boldsymbol x_i$ which is used for finding the center points and optimizing the conditional center loss. DCSL then uses an output layer parameterized with $W$ and $b$ to obtain the output vector $f(x_i)$, where $f$ is a scoring function and is consisted of the backbone and the output layer. DCSL finally applies the score-level cost matrix on the output $f(x_i)$ to obtain the new output $o_i$, which is inputted into the joint loss layer when training and into the softmax layer when testing. The joint loss is revised by

\begin{equation}
\begin{split}
&\mathcal{L}_{DCSL}  = \mathcal{L}(\xi) + \lambda_C \mathcal{L}_C \\
& = \frac{1}{m} \sum_{i=1}^{m} \log \frac{e^{-\xi w_p(W_{y_p}^{T}\boldsymbol x_i + b_{y_p})}}{\sum_{p=1}^{n} e^{\xi (W_{p}^{T}\boldsymbol x_i+b_p)}} + w_p \lambda_C\parallel \boldsymbol x_i - \boldsymbol c_{y_i}\parallel_2^2)\,.
\end{split}
\end{equation}

The workflow of learning and optimization of DCSL is shown in Algorithm~\ref{alg:Framwork}.

\section{Experiments}
\label{Experiments}
We evaluate our algorithm on a newly-collected dataset against state-of-the-art algorithms. The code and dataset will be publicly available.

\subsection{Data and Set-up}

To evaluate the performance of our method on screening of COVID-19, we collected a multi-class multi-center chest X-ray dataset. This dataset includes 2,239 examples with image-level labels. Specifically, we collected the dataset from three different sources. The first source is from a GitHub collection of chest X-rays of patients diagnosed with COVID-19\footnote{\url{https://github.com/ieee8023/covid-chestxray-dataset}.}. The second source is from a Kaggle dataset\footnote{\url{https://www.kaggle.com/andrewmvd/convid19-x-rays}}, which is thoroughly collected from the websites of 1) Radiological Society of North America (RSNA), 2) Radiopaedia, and 4) Italian Society of Medical and Interventional Radiology (SIRM). The third source is from a collection of X-ray images of bacterial and viral pneumonia~\cite{kermany2018identifying}. The collected data consists of 239 chest X-rays with confirmed COVID-19, 1,000 chest X-rays with confirmed bacterial and viral pneumonia, and 1,000 examples of healthy condition. We selected out low-quality images in the dataset to prevent unnecessary classification errors. Representative Chest X-ray images of different classes are illustrated in Figure~\ref{fig:2}, which shows the subtle differences between different classes as well as proving the necessity of fine-grained classification.

In our experiments, we conduct a three-class classification task for better verifying the proposed SCSL algorithm in the screening task. The first class is healthy X-ray images, the second class is confirmed COVID-19 X-ray images, and the third class is other confirmed pneumonia X-ray images, which include both bacterial and viral pneumonia. We employ standard five-fold cross-validation on the dataset for performance evaluation and comparison. The dataset is divided into five groups. Among them, four groups are used for training the deep neural networks, and the last group is used for testing the performance. This procedure is repeated five times until the indices of all of the subjects are obtained. The evaluation metrics include accuracy, precision, sensitivity, and F1 score. 

In order to verify the effectiveness of our proposed algorithm, we compare our designed DCSL algorithm with state-of-the-art methods: COVID-Net~\cite{wang2020covid}, VGG19~\cite{simonyan2014very}, Inceptionv3~\cite{szegedy2016rethinking}, ResNet50~\cite{he2016deep}. COVID-Net is a newly-proposed deep convolutional neural network tailored for the detection of COVID-19 from chest X-rays. VGG19, Inceptionv3, and ResNet50 are popular convolutional neural networks which have made great success in various tasks.

We implement our algorithm in Keras. We use the common VGG16 network as the backbone~\cite{simonyan2014very}. The original VGG16 network has 16 layers. There are 13 convolutional layers with a small filter with a size of $3 \times 3$ for extracting deep features. Five max-pooling layers with $2 \times 2$ kernel are deployed after each block of the convolutional layers. We set the output shape of the features of the last convolutional layer to be $7 \times 7 \times 512$ and flatten them. The original fully connected layers of VGG16 are removed and replaced by two trainable fully-connected layers. The channel numbers of the two fully-connected layers are 256 and three, respectively. Since the collected dataset is too small to obtain promising results through training the deep network from scratch, we use a transfer learning strategy, \emph{i.e.,} the parameters of the convolutional layers are initialized from the pre-trained model\footnote{\url{https://github.com/fchollet/deep-learning-models/releases}} on ImageNet \cite{deng2009imagenet}. Also, all the compared algorithms are implemented according to their open-source codes with pretraining. The chest X-ray images were resized into $224 \times 224 \times 3$. We also use an augmentation strategy to expand the dataset: each random-selected example is rotated by 15 degrees clockwise or counterclockwise. The $\lambda_C$ is set as 0.05. $\alpha$ is set to one without loss of generality. Adam optimizer is used with an initial learning rate of 1e-3. We set the training epoch to 40.

\begin{figure}[pos=th]
	\centering
	\includegraphics[width=0.25\textwidth]{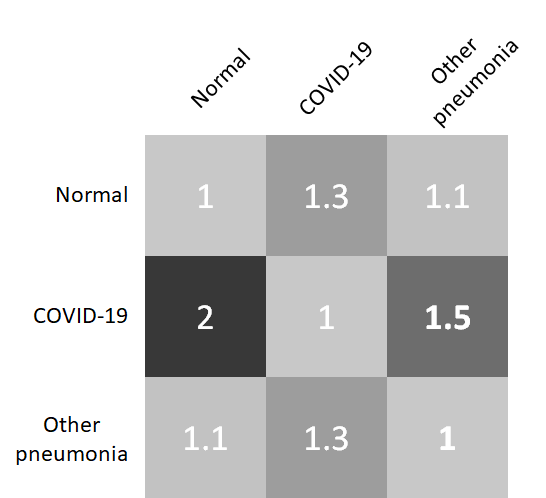}
	\caption{\label{fig:cost_matrix} An illustration of the score-level cost matrix designed according to the clinical expert experience.}
\end{figure}

The score-level cost matrix is designed according to the clinical expert experience as following. First of all, the cost of misclassifying COVID-19 is higher than misclassifying other classes. Among them, the cost of misclassifying COVID-19 into healthy patients is higher than the cost of misclassifying COVID-19 into other pneumonia patients. Second, the cost of misclassifying other pneumonia is smaller than the cost of misclassifying COVID-19. Specifically, the cost of misclassifying other pneumonia into COVID-19 is higher than the cost of misclassifying other pneumonia into healthy patients. Third, the cost of misclassification of healthy people is smaller than in the previous two situations. Among them, the cost of misclassifying a healthy person into COVID-19 is greater than the cost of misclassifying a healthy person into other pneumonia. Accordingly, the final score-level cost matrix is designed as illustrated in Figure~\ref{fig:cost_matrix}.

\begin{table}[pos=ht]
	\scriptsize
	\centering
	\caption{Classification results of multiple algorithms on three classes: normal, COVID-19, and other pneumonia.}
	\begin{tabular}{l|p{0.98cm}<{\centering} p{0.98cm}<{\centering} p{0.98cm}<{\centering} p{0.98cm}<{\centering}}
		\toprule
		Method & Accuracy & Precision & Sensitivity & F1-score \\
		\midrule
		VGG19 & 0.9196 & 0.9238 & 0.9196 & 0.9200 \\
		Inceptionv3 & 0.9107 & 0.9113 & 0.9107 & 0.9105 \\
		ResNet50 & 0.9107 & 0.9135 & 0.9107 & 0.9104 \\
		COVID-Net & 0.9330 & 0.9339 & 0.9330 & 0.9332 \\
		DCSL (Ours) & \textbf {0.9701} & \textbf {0.9700} & \textbf {0.9709} & \textbf {0.9698} \\ 
		\bottomrule
	\end{tabular}
	\label{tab1}
\end{table}

\begin{figure}[pos=ht]
	\centering
	\subfigure[DCSL]{\includegraphics[width=40mm]{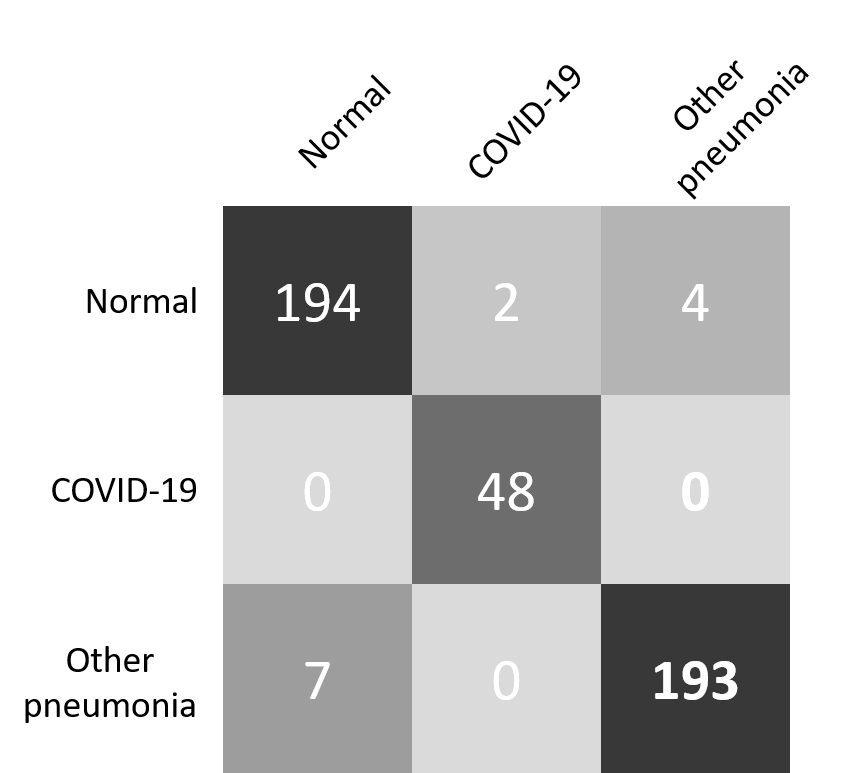}}
	\subfigure[COVID-Net]{\includegraphics[width=40mm]{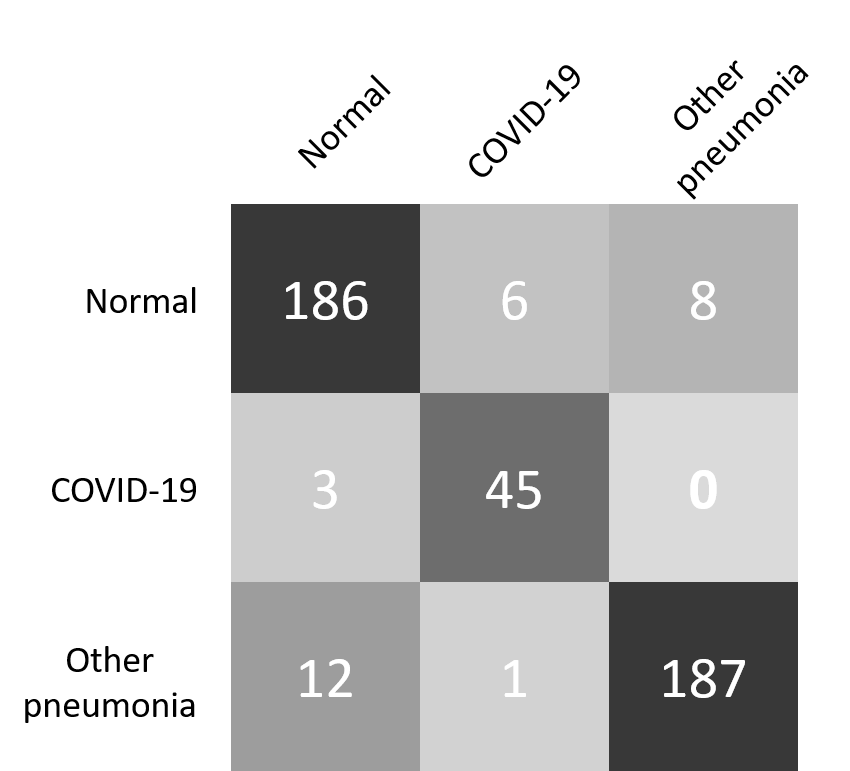}}
	\caption{The confusion matrixes of DCSL and COVID-Net.}
	\label{cm_bc}
\end{figure}

\begin{figure}[pos=ht]
	\centering
	\includegraphics[width=0.45\textwidth]{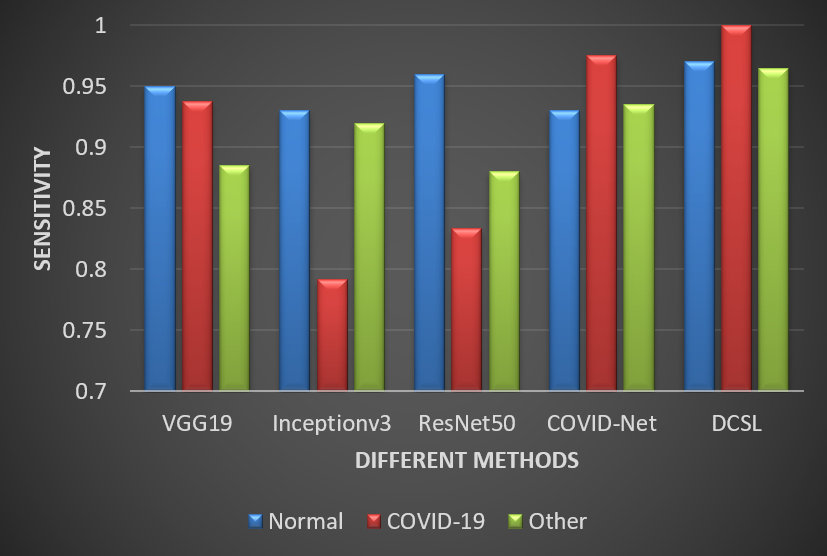}
	\caption{An illustration of the sensitivities values of each class.}
	\label{sensitivity}
\end{figure}

\begin{figure*}
	\begin{minipage}[t]{0.248\linewidth} 
		\centering
		\includegraphics[width=0.8\textwidth, height=0.7\textwidth]{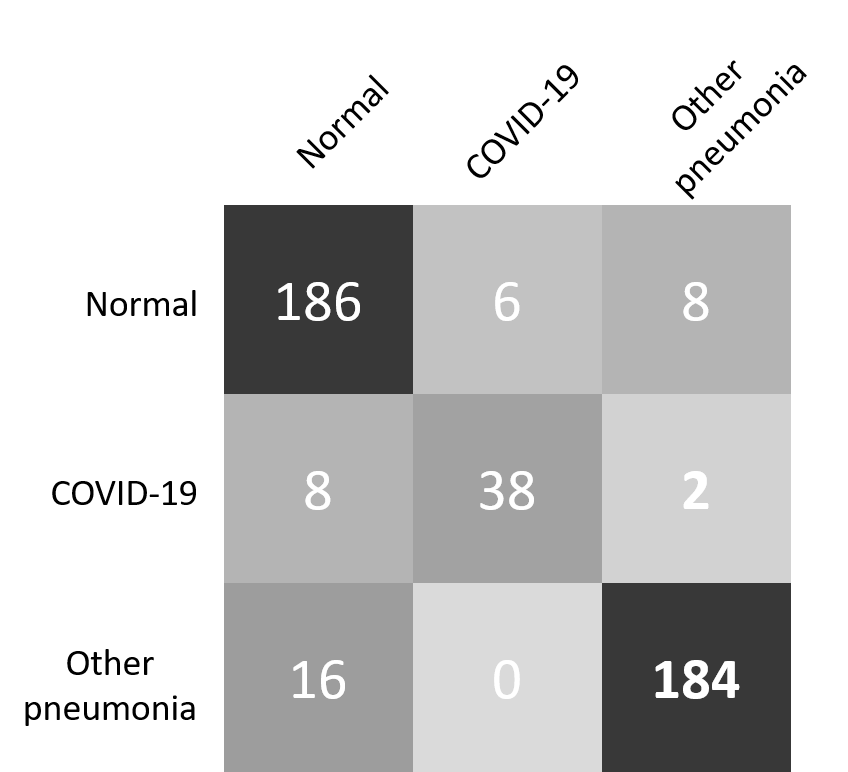}
		\centerline{(a) Softmax Loss}
		\label{fig3:side:a}
	\end{minipage}%
	\begin{minipage}[t]{0.248\linewidth}
		\centering
		\includegraphics[width=0.8\textwidth, height=0.7\textwidth]{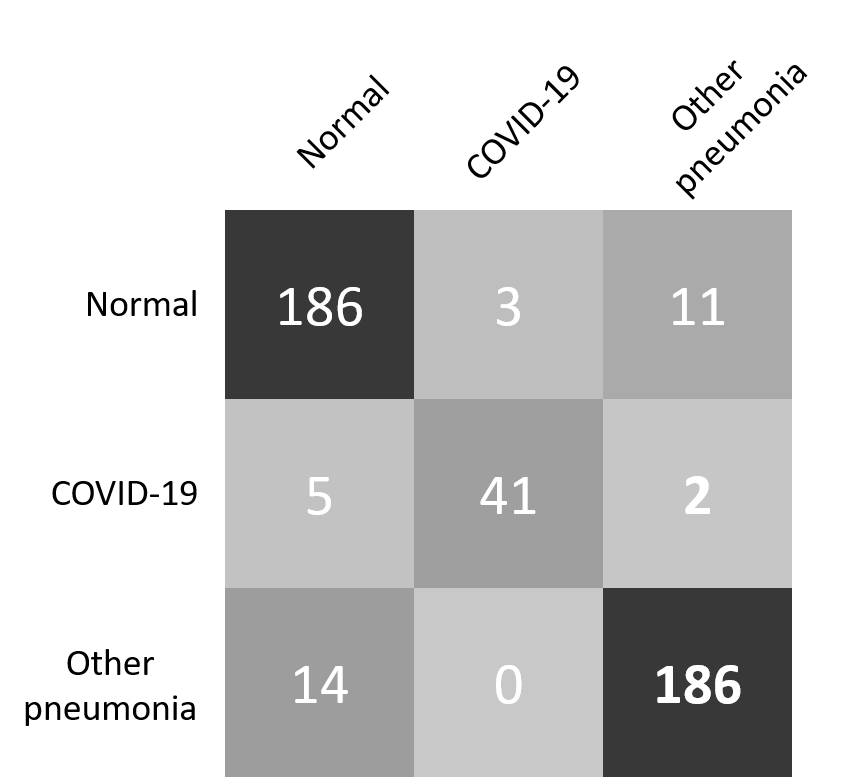}
		\centerline{(b) Softmax+Center Loss}
		\label{fig3:side:b}
	\end{minipage}
	\begin{minipage}[t]{0.248\linewidth} 
		\centering
		\includegraphics[width=0.8\textwidth, height=0.7\textwidth]{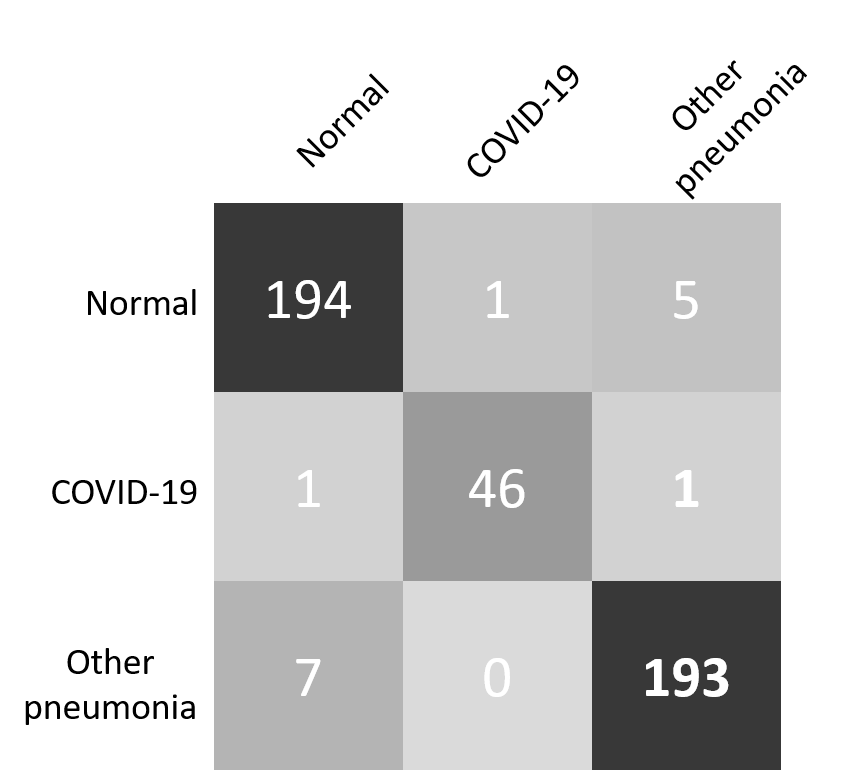}
		\centerline{(c) Softmax Loss + CCL}
		\label{fig3:side:c}
	\end{minipage}%
	\begin{minipage}[t]{0.248\linewidth}
		\centering
		\includegraphics[width=0.8\textwidth, height=0.7\textwidth]{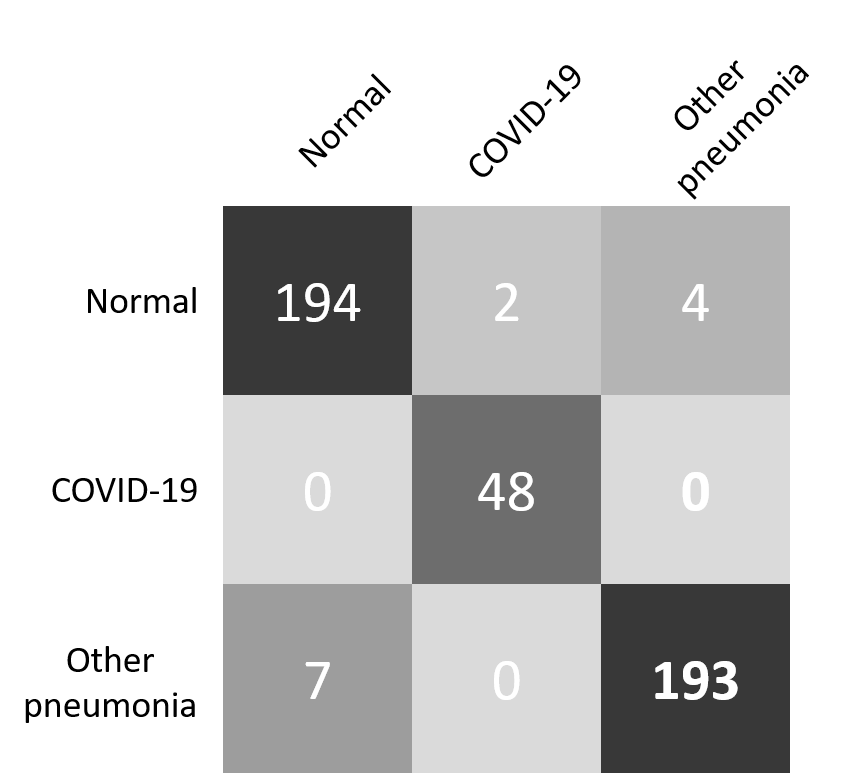}
		\centerline{(d) DCSL}
		\label{fig3:side:d}
	\end{minipage}
	\caption{\label{fig:3} The confusion matrixes of ablation studies from \textbf{Softmax Loss}, \textbf{Softmax+Center Loss}, \textbf{Softmax Loss + Conditional Center Loss (CCL)}, and \textbf{DCSL}. Our algorithm shows a higher and balance performance.}
\end{figure*}

\begin{table}[pos=ht]
	\scriptsize
	\centering
	\caption{This table shows the results of the ablation study of DCSL on the accuracy, precision, sensitivity, and F1-score.}
	\begin{tabular}{l|p{0.98cm}<{\centering} p{0.98cm}<{\centering} p{0.98cm}<{\centering} p{0.98cm}<{\centering}}
		\toprule
		Method & Accuracy & Precision & Sensitivity & F1-score \\
		\midrule
		Softmax Loss & 0.9163 & 0.9173 & 0.9162 & 0.9158 \\
		Softmax Loss+CL & 0.9208 & 0.9229 & 0.9207 & 0.9208 \\
		Softmax Loss+CCL & \textbf{0.9721} & \textbf{0.9723} & \textbf{0.9721} & \textbf{0.9721} \\ 
		\bottomrule
	\end{tabular}
	\label{tab:widgets2}
\end{table}

\subsection{Results}

The proposed discriminative cost-sensitive learning algorithm (DCSL) achieves the highest results on the screening of COVID-19 from chest X-rays. Table~\ref{tab1} reports the results of our algorithm and compared algorithms. Our algorithm obtains a classification accuracy of 97.1\%, a precision of 97.0\%, a sensitivity of 97.09\%, and an F1-score of 96.98\%. Our algorithm remarkably outperforms COVID-Net~\cite{wang2020covid}, which achieves state-of-the-art results before our work. Also, our algorithm significantly outperforms all the compared algorithms on all the metrics. As shown in Figure~\ref{fig:2}, even both the complex lung structures and indiscernible infection areas lead to unusual difficulties; our algorithm still obtains accurate performance, which demonstrates its robust strengths.

Figure~\ref{cm_bc} displays the confusion matrixes of our algorithm and COVID-Net. Owing to our score-level cost-sensitive learning, we achieve 100\% accuracy in the class of COVID-19. Such a result demonstrates the effectiveness of incorporating the score-level matrix after the output layer of deep neural networks to modify the learning process. Figure~\ref{sensitivity} presents the sensitivities of each class, where our algorithm achieves 100\% sensitivity of COVID-19, which is much higher than compared methods. These results once verify the advantages of score-level cost-sensitive learning. Both Figure~\ref{cm_bc} and Figure~\ref{sensitivity} show that our algorithm also achieves the highest accuracy in other classes, which demonstrates the critical role of conditional center loss that can improve the intra-class compactness evenly.

We further perform statistical analysis to ensure that the experimental results have statistical significance. A paired t-test between the COVID-Net and our algorithm is at a 5\% significance level with a p-value of 0.010. This analysis result clearly shows that the improvement of our method is noticeable. The p-values of the VGG19, Inceptionv3, and ResNet50 models are less than 0.05, which proves that popular classifiers are not suitable for the task of screening COVID-19 from chest X-rays. These analyses verify that our insight that viewing the screening of COVID-19 from chest X-rays as a fine-grained cost-sensitive classification task is correct.

\begin{figure}[pos=t]
	\begin{minipage}[t]{0.497\linewidth} 
		\centering
		\includegraphics[width=0.9\textwidth]{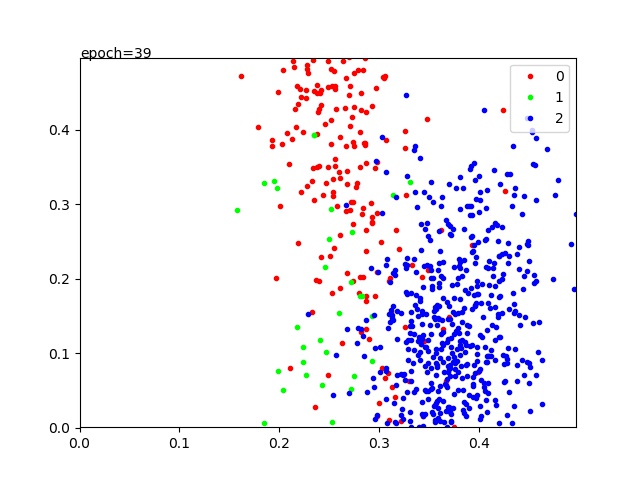}
		\centerline{(a) Softmax Loss}
		\label{fig5:side:a}
	\end{minipage}%
	\begin{minipage}[t]{0.497\linewidth}
		\centering
		\includegraphics[width=0.9\textwidth]{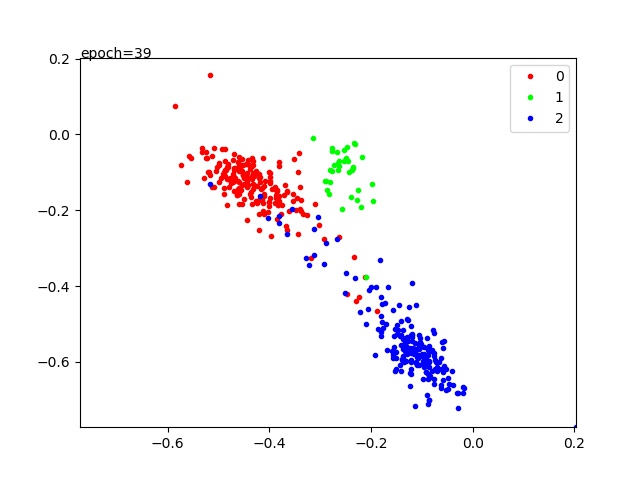}
		\centerline{(b) Softmax Loss + CCL}
		\label{fig5:side:b}
	\end{minipage}
	\caption{The t-SNE visualization of deep features from the training data. The feature distribution of using \textbf{Softmax Loss + Conditional Center Loss (CCL)} are more discriminative than using \textbf{Softmax Loss} only. The points of different colors denote the deep features from different classes (0: normal; 1: COVID-19; 2: other pneumonia).}
	\label{tsne}
\end{figure}

\subsection{Analysis}

This section further gives in-depth ablation studies to demonstrate the effect of conditional center loss (CCL) and score-level cost-sensitive learning (SLCSL), respectively. We construct four ablation models based on the backbone network VGG-16. The first model is only using the cross-entropy loss with softmax called as \textbf{Softmax Loss}. The second model combines center loss and cross-entropy loss with softmax called as \textbf{Softmax+Center Loss}. Similarly, the third model combines conditional center loss and softmax loss called as \textbf{Softmax Loss + CCL}. The final model is our algorithm \textbf{DCSL} that combines score-level cost-sensitive learning, conditional center loss, and softmax loss.

\begin{figure}[pos=ht]
	\setlength{\abovecaptionskip}{-0.1cm}   
	\setlength{\belowcaptionskip}{-1cm}   
	\centering
	\subfigure[Softmax Loss]{\includegraphics[width=29.1mm]{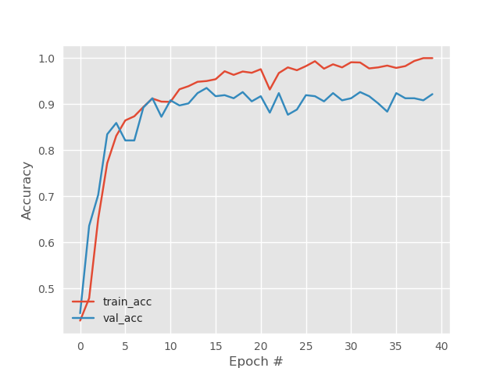}}
	\hspace{-4.mm}
	\subfigure[Softmax+Center Loss]{\includegraphics[width=29.1mm]{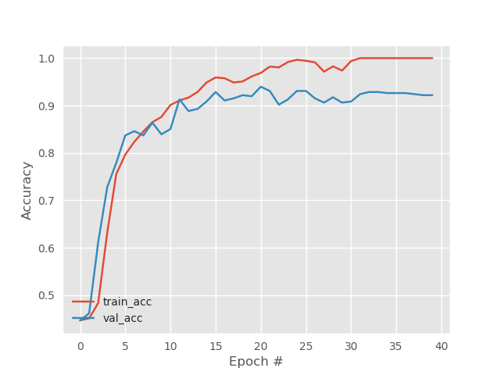}}
	\hspace{-4.mm}
	\subfigure[DCSL]{\includegraphics[width=29.1mm]{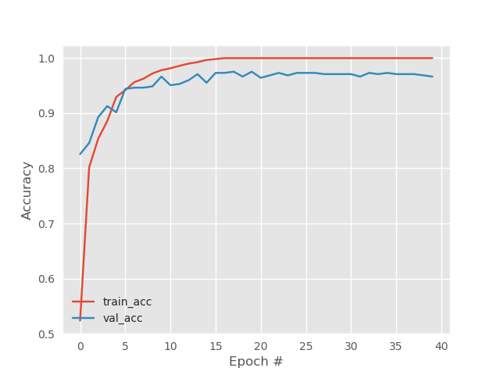}}
	\caption{Accuracy curves in the training and validation period. They show that DCSL has higher accuracy with stable convergence.}
	\label{fig:7}
\end{figure}

\begin{figure}[pos=ht]
	\setlength{\abovecaptionskip}{-0.1cm}   
	\setlength{\belowcaptionskip}{-1cm}   
	\centering
	\subfigure[Softmax Loss]{\includegraphics[width=29.1mm]{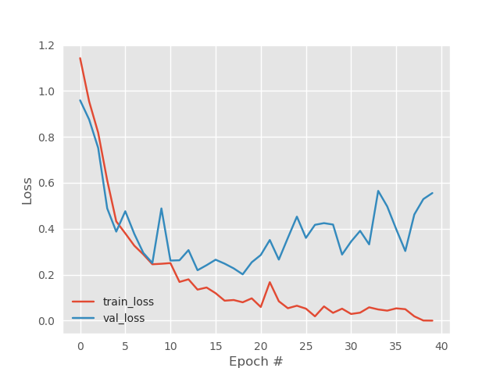}}
	\hspace{-4.mm}
	\subfigure[Softmax+Center Loss]{\includegraphics[width=29.1mm]{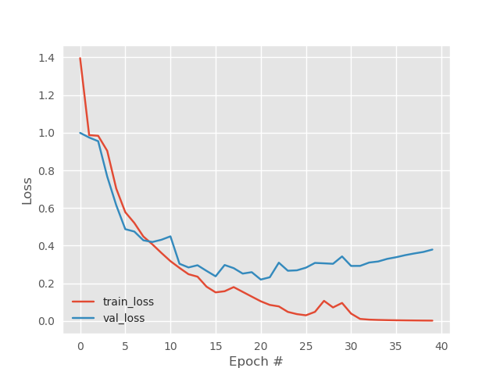}}
	\hspace{-4.mm}
	\subfigure[DCSL]{\includegraphics[width=29.1mm]{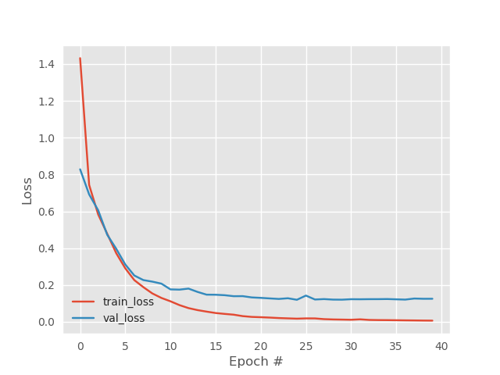}}
	\caption{Loss curves in the training and validation period. It can be observed that DCSL has great stability and fast convergence.}
	\label{fig:6}
\end{figure}

Generally speaking, our final model DCSL achieves the best performance than the other ablation models, as shown in Figure~\ref{fig:3}. These confusion matrixes strongly prove that our algorithm can accurately screen COVID-19 from chest X-rays without any missing case. Both Figure~\ref{fig:7} and Figure~\ref{fig:6} demonstrate the convergence and stability of DCSL in the training and validation period. These excellent results show that our algorithm successfully achieves accurate and robust screening of COVID-19 from chest X-rays. These extensive results once verify the correctness of our insight that this task is a fine-grained cost-sensitive classification problem.


\subsubsection{Conditional Center Loss}
Table~\ref{tab:widgets2} reports the results of out ablation study on different loss functions. Our conditional center loss (Softmax Loss + CCL) remarkably outperforms the center loss and softmax loss. These results demonstrate the importance of considering the class-conditional information when updating the center points and optimizing the center loss. Figure~\ref{tsne} presents the t-SNE visualization of deep features from the training data. It verifies that the conditional center loss can contribute to improving the intra-class compactness.

Moreover, Figure~\ref{fig:3} shows that the conditional center loss has fewer mistakes and achieves a balance performance on the three classes. Also, Both Figure~\ref{fig:7} and Figure~\ref{fig:6} show that the conditional center loss has excellent stability and fast convergence. In summary, the conditional center loss has a significant impact on the performance of our proposed architecture. When the conditional center loss is not used, the result of classification is obviously decreased, and the learned deep features contain significant intra-class variations.

\begin{figure}[pos=ht]
	\centering
	\includegraphics[width=0.48\textwidth]{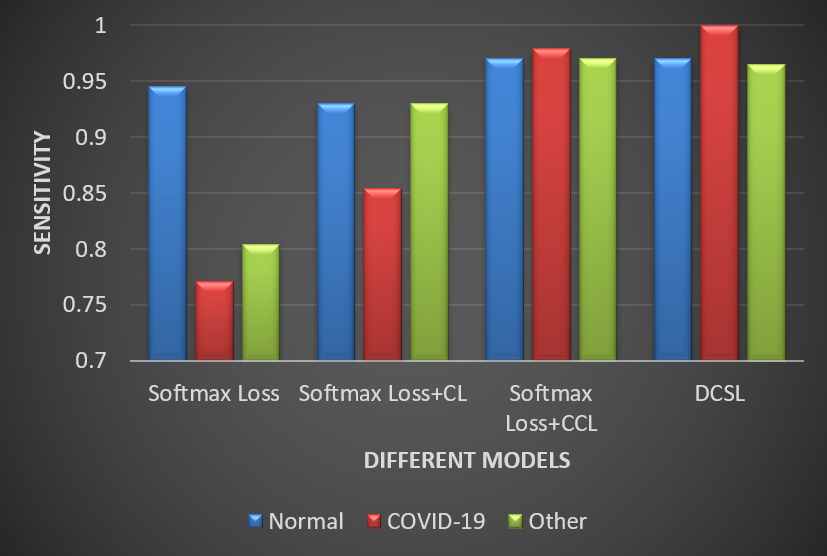}
	\caption{Classification results of the ablation studies from \textbf{Softmax Loss}, \textbf{Softmax+Center Loss (CL)}, \textbf{Softmax Loss + Conditional Center Loss (CCL)}, and \textbf{DCSL}. Owning to the score-level cost-sensitive learning, we achieve the highest sensitivity on COVID-19 class. The conditional center loss also plays a vital role in the improvement of performance. These results demonstrate the effectiveness of DCSL.}
	\label{fig:8}
\end{figure}

\subsubsection{Score-Level Cost-Sensitive Learning}
Another goal of this work is to enhance the sensitivity of COVID-19 without decreasing the overall classification accuracy. Although all the results have verified the advantages of score-level cost-sensitive learning, we should dissect its strengths. Figure~\ref{fig:3} shows that using score-level cost-sensitive learning achieves zeros mistake of the COVID-19 class. Moreover, Figure~\ref{fig:8} demonstrates that DCSL makes the 100\% sensitivity in COVID-19 without decreasing the overall classification accuracy. Experimental results show that DCSL can significantly improve the sensitivity and precision of COVID-19.

To conclude, cost-sensitive learning plays a crucial role in the screening of  COVID-19. During the global outbreak of COVID-19, the cost of misclassifying COVID-19 patients into other types of pneumonia or even healthy people are much higher than the cost of misclassifying other classes. The proposed score-level cost-sensitive learning has significantly improved the sensitivity of COVID-19, proving our hypothesis that cost-sensitive learning is very suitable for the new task.

\section{Conclusion}
\label{Conclusion}

In this paper, we reported a new attempt for the fine-grained cost-sensitive screening of COVID-19 from chest X-rays. We proposed a novel discriminative cost-sensitive learning (DCSL) that includes a conditional center loss function and a score-level cost-sensitive learning module. To the best of our knowledge, this is the first method that formulates this novel application as a fine-grained cost-sensitive classification problem. Extensive results have demonstrated that DCSL can achieve reliable and accurate results. In-depth analyses have revealed the effectiveness and potential of DCSL as a clinical tool to relieve radiologists from laborious workloads, such that contribute to the quickly large-scale screening of COVID-19.


\section*{Acknowledgment}
This work was partly funded by Natural Science Foundation of China (No.61872225); Introduction and Cultivation Program for Young Creative Talents in Colleges and Universities of Shandong Province (No.173); the Natural Science Foundation of Shandong Province (No.ZR2019ZD04, No.ZR2015FM010); the Project of Science and technology plan of Shandong Higher Education Institutions Program (N
o.J15LN20); the Project of Shandong Province Medical and Health Technology Development Program (No.2016WS0577).

\bibliographystyle{cas-model2-names}
\bibliography{mybibliography}

\end{document}